\documentclass{elsart}[10pt]

\usepackage{graphicx}

\newcommand{\snm}{\sqrt{s_{\rm NN}}}

\newcommand{\sne}[1]{$\snm = #1$}

\newcommand{\spcmfull}{\langle \rho_{\rm SC} / \epsilon_0 \rangle}
\newcommand{\spcm}{\rho_{\rm SC}}
\newcommand{\spc}{$\spcm$}
\newcommand{\spcrm}{\spcm^{\rm sim}}
\newcommand{\spcum}{\spcm^{\rm used}}
\newcommand{\spcnm}{\spcm^{\rm new}}
\newcommand{\spcom}{\spcm^{\rm obs}}
\newcommand{\spco}{$\spcom$}
\newcommand{\spcop}{$\spcm^{\rm prepass}$}
\newcommand{\glm}{\rho_{\rm leak}}
\newcommand{\gl}{$\glm$}
\newcommand{\glrm}{\glm^{\rm sim}}
\newcommand{\glom}{\glm^{\rm obs}}

\newcommand{\ctrkm}{C_{\rm track}^{\rm sim}}
\newcommand{\ctrk}{$\ctrkm$}

\newcommand{\dcam}{{\rm sDCA}_{\rm track}}
\newcommand{\dcaom}{\dcam^{\rm obs}}
\newcommand{\dcarm}{\dcam^{\rm sim}}
\newcommand{\dcae}{$\langle$sDCA$\rangle$}

\begin{document}

\begin{frontmatter}



\title{Correcting for Distortions due to Ionization in the STAR TPC}

\author[BNL]{G. Van Buren\corauthref{cor}},
\corauth[cor]{Corresponding author. Address: BNL, Bldg. 510A,
Upton, NY 11973-5000, USA. Tel: 631-344-7953. Fax:
631-344-4206. Email: gene@bnl.gov.}
\author[BNL]{L. Didenko},
\author[BNL]{J. Dunlop},
\author[BNL]{Y. Fisyak},
\author[BNL]{J. Lauret},
\author[BNL]{A. Lebedev},
\author[Purdue]{B. Stringfellow},
\author[LBNL]{J.H. Thomas},
\author[LBNL]{H. Wieman}

\address[BNL]{Brookhaven National Laboratory, Upton, NY 11973, USA}
\address[LBNL]{Lawrence Berkeley National Laboratory, Berkeley, CA 94720, USA}
\address[Purdue]{Purdue University, West Lafayette, IN 47907, USA}

\begin{abstract}

Physics goals of the STAR Experiment at RHIC in recent (and future) years drive the need to operate the STAR TPC at ever higher luminosities, leading to increased ionization levels in the TPC gas. The resulting ionic space charge introduces field distortions in the detector which impact tracking performance. Further complications arise from ionic charge leakage into the main TPC volume from the high gain anode region. STAR has implemented corrections for these distortions based on measures of luminosity, which we present here. Additionally, we highlight a novel approach to applying the corrections on an event-by-event basis applicable in conditions of rapidly varying ionization sources.
\end{abstract}

\end{frontmatter}

\section{Introduction}
\label{sec:intro}

The time projection chamber (TPC) used by the STAR experiment
at RHIC has several potential sources of field distortions~\cite{TPC}.
While most of these sources are static, the buildup
of slow-drifting positively charged ions in the volume gas
generated from standard operation of the TPC
varies with the quantity of charged particles traversing the TPC, and thereby both
the luminosity of the collider and the multiplicity of charged
particles emitted by the collisions. The variations in this
``space charge'' can occur on
time scales down to what it takes the ions to drift the length
of the chamber, which is approximately one half second for the STAR TPC.

\section{Space Charge Distortions}
\label{sec:dist}

Modeling the distortions due to space charge is a straightforward
process beginning with a postulation of the typical three-dimensional
distribution of ionization in the TPC. The nearest measure we
have of this in STAR is a record of the distribution of electron
clusters reaching the TPC endcap averaged over many events
using a so-called ``zero-bias'' trigger (which is random with respect
to collision times, removing any biases related to the definition
of a collision). This measure integrates out any drift-direction
dependencies, but compares well in radial dependence (approximately
as inverse radius squared) for
\sne{200} Au+Au collisions to a simulation using the
HIJET event generator~\cite{HIJET}.
The simulation indicates a uniform distribution of charge in
the drift direction.

We use the HIJET charge distribution integrated along
the distance from the endcap to any point in space
(representing the effect of continual collision contributions)
in conjunction with the
boundary conditions of grounded
surfaces surrounding the TPC gas volume to solve for the
electrical potential due to space charge. An analytical
solution is not achievable, so we use a numerical
relaxation to solve for the potential
on a grid in two dimensions
(with assumed azimuthal symmetry) and interpolate.
An electric field is obtained from the
potential and is treated as a perturbation atop the
normal drift field. The distortions to the measured positions of
electron clusters are then calculated by integrating
the effects of this perturbing field (which depend
on operating conditions of the chamber) along the path from
a point in the TPC to the endcap where the clusters
are measured~\cite{TPC}. The amplitude of this distortion
is directly proportional to the quantity of space
charge (\spc) present. In practice, we calibrate the
average charge density over the volume of the chamber: $\spcmfull$.

\begin{figure}
\begin{center}
\includegraphics*[width=67mm]{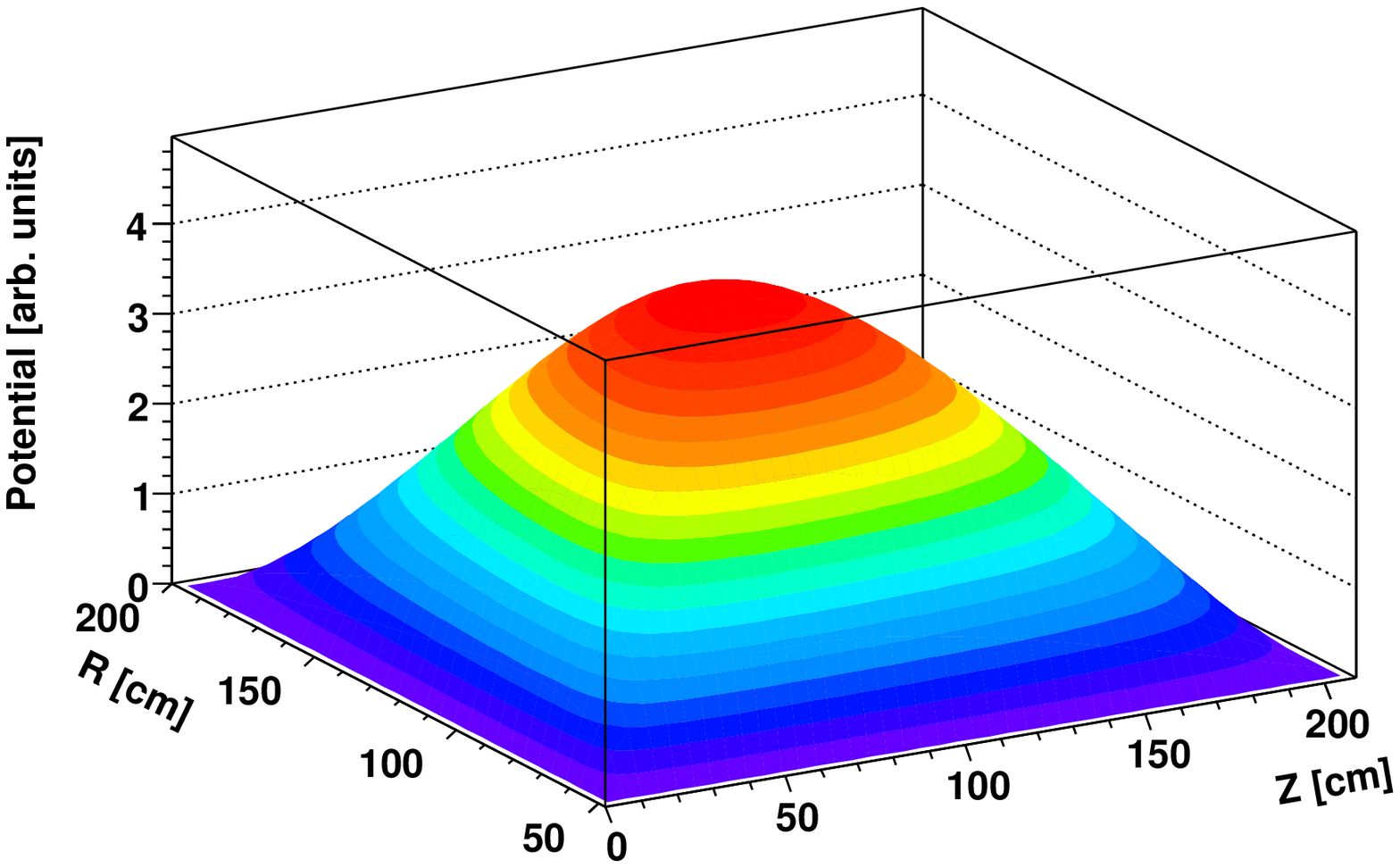}
\includegraphics*[width=67mm]{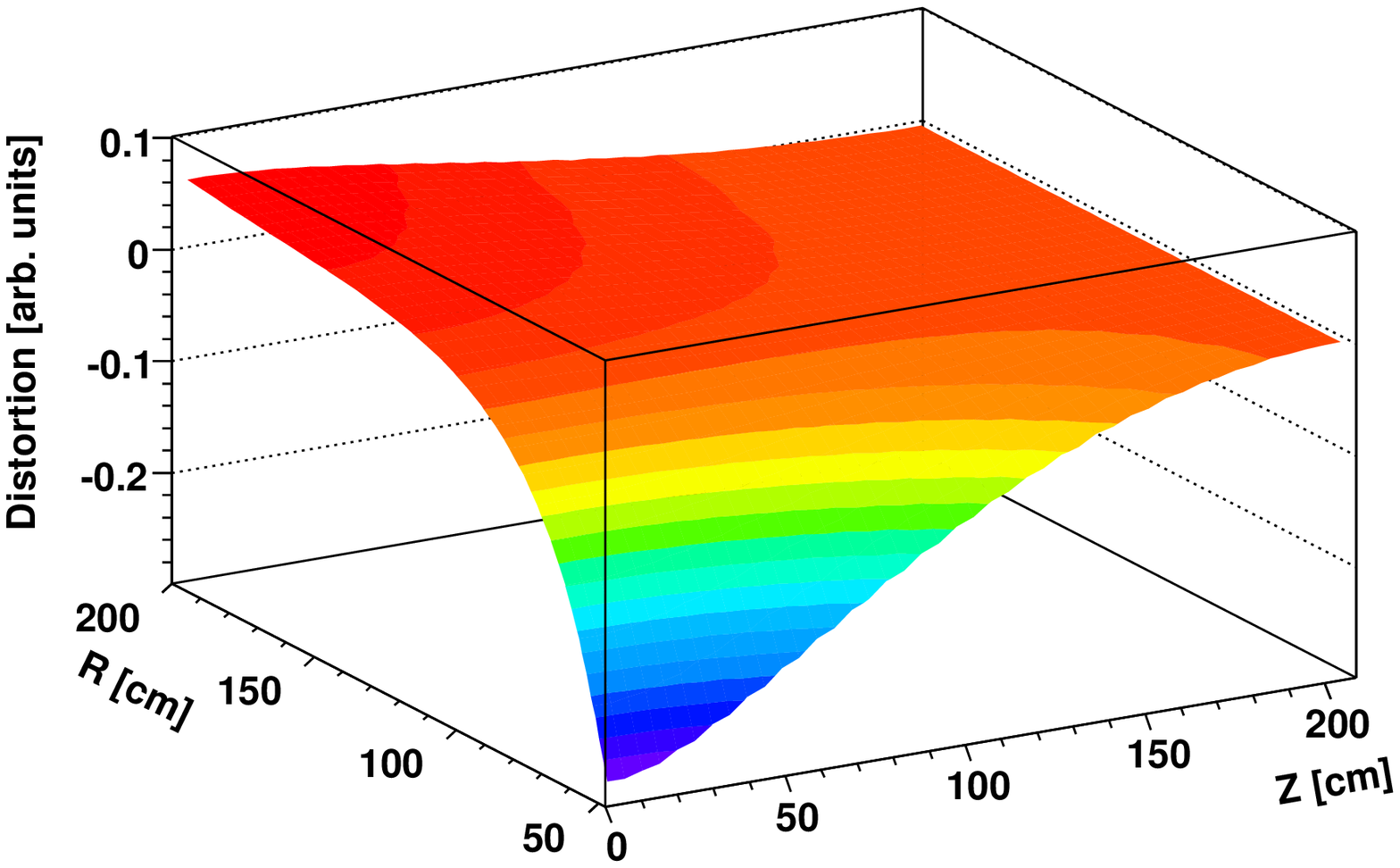}
\end{center}
\vskip -0.5cm
\caption{ \small
Simulated shape of the potential due to space charge in the TPC (left) and
the azimuthal distortions of electron clusters (right)
caused by drifting through that potential as a function of radius R and drift Z.
The cathode is at Z=0, and electron clusters drift to
the endcaps at high Z.
} \label{fig:SC}
\end{figure}

Because the Lorentz force on the drifting electron
clusters is proportional to the cross product of
the electric and magnetic field
(aligned along the drift direction in STAR) vectors, 
the principal distortion of consequence is azimuthal,
and is plotted in Fig.~\ref{fig:SC}.
This distortion has the effect of rotating
reconstructed tracks in the transverse plane about a
point midway along their path through the TPC.

\section{Space Charge Corrections}
\label{sec:cor}

Knowing \spc~is sufficient to subtract the
calculated distortions from measured electron
cluster positions to obtain their approximate
original, undistorted positions.
In the absence of direct measures of \spc,
a measure of the distortion to tracks
(fit from distorted clusters)
may suffice to indirectly determine \spc.
Simulation shows that for any given distorted primary particle track,
its signed distance of closest
approach (sDCA)\footnote{The sign is determined
by the Z-direction of the cross product of the track momentum
vector at its closest approach and the vector pointing to the
collision vertex, essentially identifying on which ``side'' of
the vertex the reconstructed track passes.} to the
collision vertex is approximately linearly proportional
to space charge, and we can obtain
$\ctrkm = \spcrm / \dcarm$, where \ctrk~depends
on the locations of points on the track.
Each real track can then
be used to derive an observed space charge:
\[
\spcom = \ctrkm \cdot \dcaom =  \spcrm \cdot ( \dcaom / \dcarm )
\]
To understand the scale of this distortion, it is
worthwhile to note that some recorded events exhibited
beyond 1cm offsets in \dcae, the mean of their
$\dcaom$~distributions.\footnote{We use only TPC tracks
with at least 25 points, pseudorapidity within $\pm$1, and
transverse momentum between 0.3-2.0 GeV/c for
all sDCA and \spc~measurements.}

\begin{figure}
\begin{center}
\includegraphics*[width=67mm]{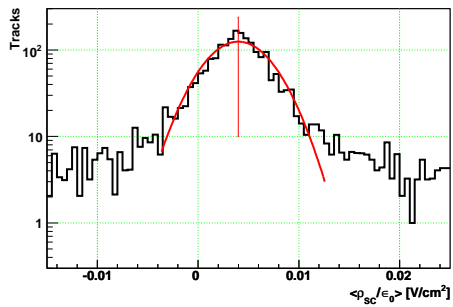}
\end{center}
\vskip -0.3cm
\caption{ \small
Observed space charge density
(averaged over the volume of the TPC) determined
from individual tracks in a single high-multiplicity event.
The mean of a Gaussian peak (formed from primaries) is fit to
extract \spco~for that event.
} \label{fig:single}
\end{figure}

A distribution of \spco~values from any given collision
event will include a background from secondaries which
naturally do not point to the collision vertex, and will be
smeared by the intrinsic resolution of the TPC to
measure sDCA. As seen in
the distribution from a single very high
multiplicity event in Fig.~\ref{fig:single}, 
the centroid of a peak formed by primaries provides
a means to determine \spco~more accurately.

To be effective, the value of \spc~used to correct the distortions
must be updated on time scales shorter than the
fluctuations caused by collider operating conditions.
During the 2000 through 2003
years of operating RHIC, scalers of trigger counter rates
recorded online
(during runs) every 30
seconds served to measure these
fluctuations sufficiently~\cite{trig}.
Along with a significant
luminosity increase in 2004, however, these
fluctuations were observed in the systematic
behavior of sDCA distributions on sub-second time scales.

An event-by-event (E-by-E) method
using only \spco~from individually recorded events suffers
from insufficient statistics to get a good measure in most
events. To compensate, we can take advantage of
the fact that \spc~fluctuations cannot occur on
time scales much shorter than the drift time of
ions in the TPC. We do this by building a running
sum of \spco~from each event and previous
events downweighted appropriately by their age. Because we
measure \spco~from events which have already
been corrected with some value $\spcum$,
we set the new value to be
$\spcnm = \spcum + \spcom$. This method is
self-correcting in that even if the conversion
factors \ctrk~are not perfect,
\spc~will quickly converge to a value which brings the
sDCA distributions to peak at zero.

Weaknesses in this technique include events at the
start of data files (for which there are no previous
events), sizable time gaps
between some events, and series of low multiplicity
events for which insufficient statistics are obtained
within short time scales. The first problem is solved by
performing a prepass on the first few events in
each file to determine a viable initial \spcop,
which is then used in the production pass until the
E-by-E method becomes applicable.
The latter issues are
handled by falling back to
\spcop~for such events until
the E-by-E method can again be useful.
Backgrounds which introduce charge
distributions different from the HIJET model
can also degrade performance.

\begin{figure}
\begin{center}
\includegraphics*[width=67mm]{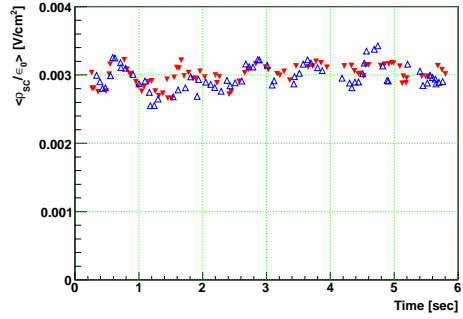}
\end{center}
\vskip -0.3cm
\caption{ \small
Observed volume-averaged space charge density measured and
used in the E-by-E method for two selections
of independent but concurrent events
versus time.
} \label{fig:qa}
\end{figure}

Fig.~\ref{fig:qa} demonstrates that the fluctuations
in \spc~determined by the E-by-E method
are not artificial. In two independent but concurrent sets of
events, similar behaviors can be seen on sub-second
time scales, while differences
illustrate the uncertainty on $\spcmfull$~in the method
of about 0.0001 V/cm$^2$.

\section{Ion Leakage Around the Gated Grid}
\label{sec:grid}

\begin{figure}[b]
\begin{center}
\includegraphics*[width=67mm]{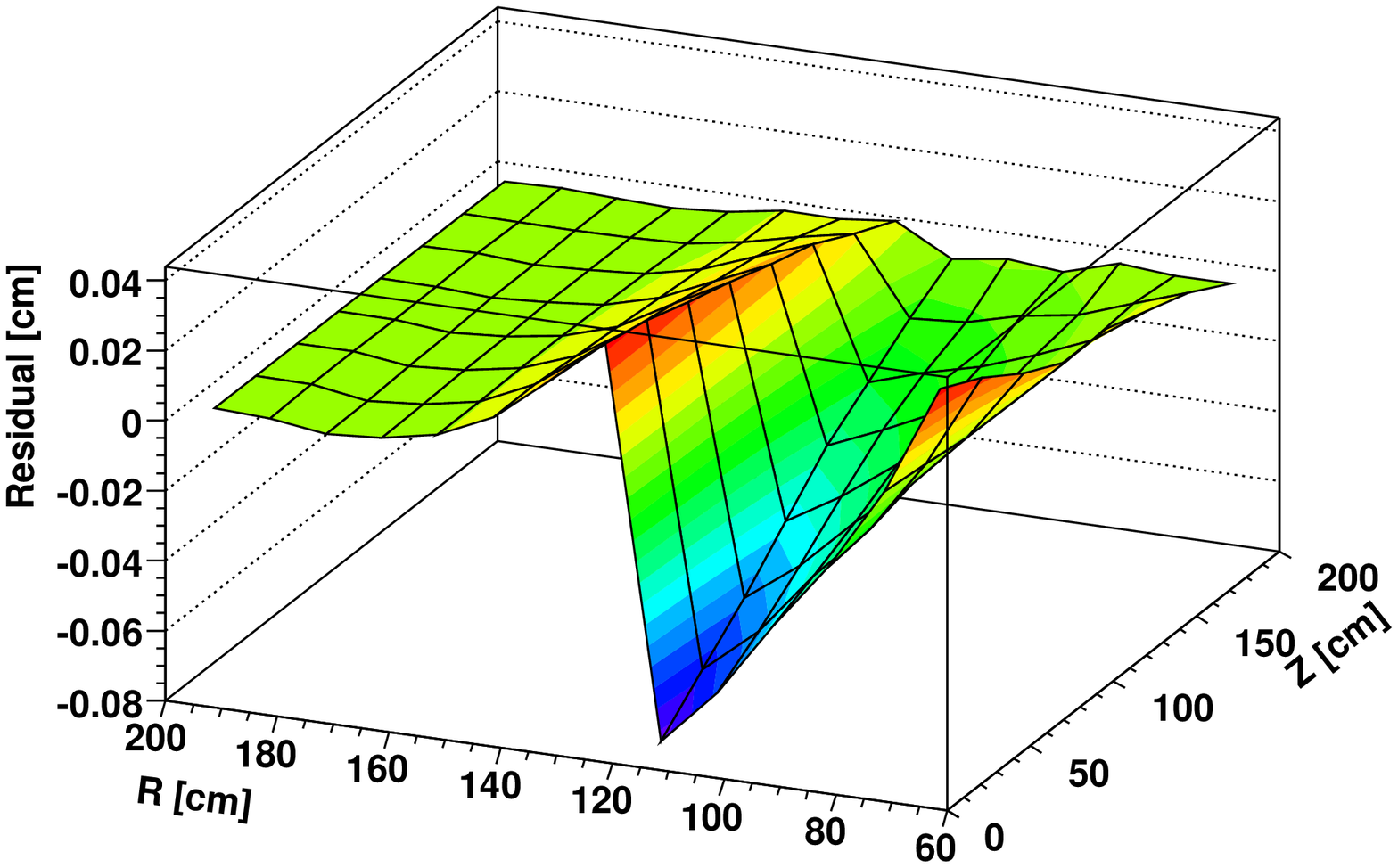}
\includegraphics*[width=67mm]{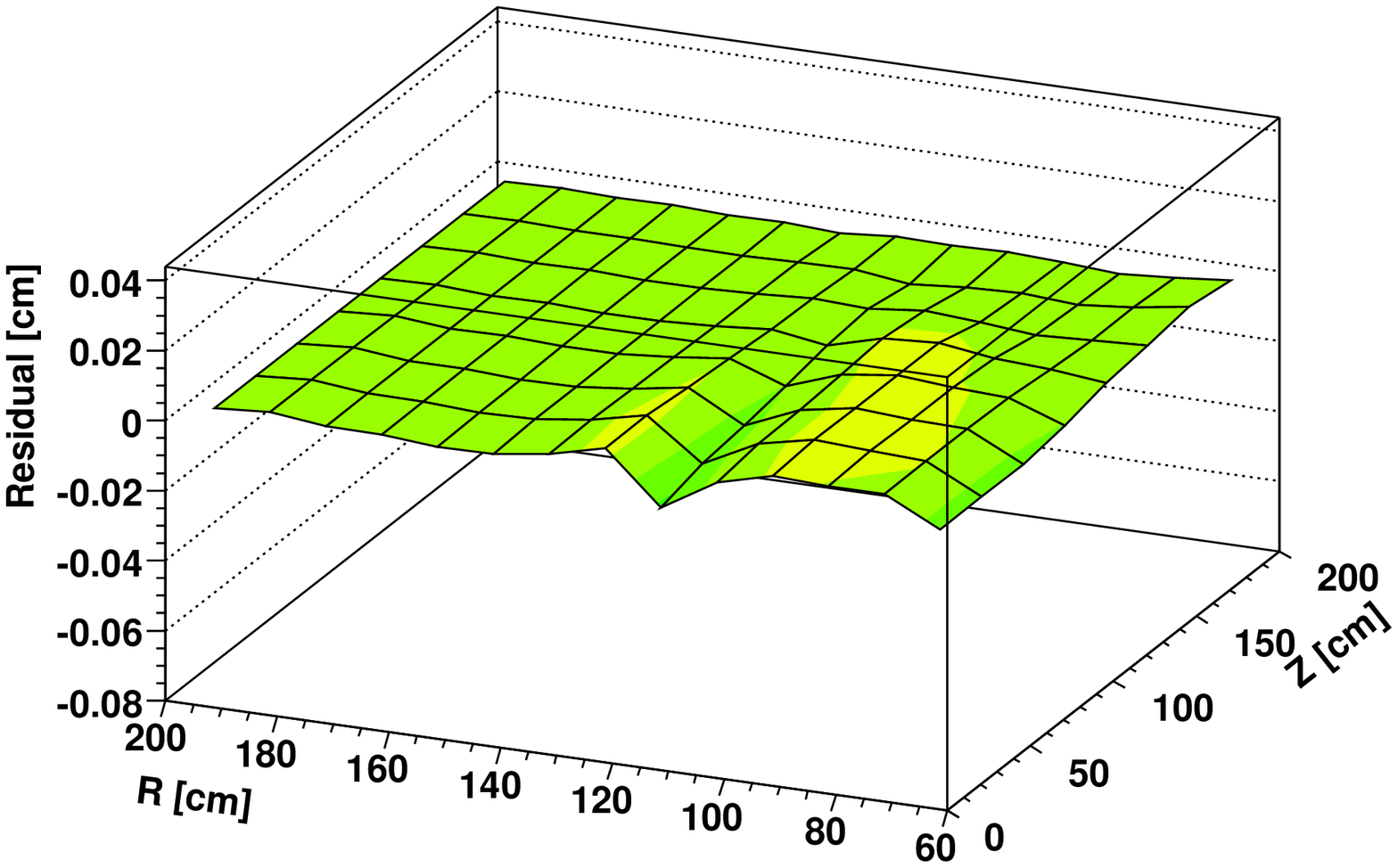}
\end{center}
\vskip -0.4cm
\caption{ \small
Residuals of TPC tracks over R and Z
in a selection of events acquired during high luminosity
before (left) and after (right) leakage distortion corrections.
The gap between sector wire chambers
is at R $\approx$ 122cm.
} \label{fig:gl}
\end{figure}

Studying residuals of TPC cluster positions from
track fits revealed that an additional
source of ions is also present in the TPC. 
A discontinuity in the residuals at the gap between
the inner and outer readout wire chambers of the
TPC, evident in Fig.~\ref{fig:gl}, is consistent with
incomplete blockage (by the gated grid) at this gap
of ions created in the
high gain region around the anode wires.
This allows a sheet of ions to flow from this gap
across the TPC gas volume
to the cathode.

Again, we can model the distortions from this leak
around the gated grid in the same manner as the
space charge, providing a map of cluster position corrections
whose magnitude is proportional to the amount
of leaked charge (\gl). These distortions
similarly affect sDCA, and \gl~was found to
scale with collision rates in the same manner
as \spc. A calibration was performed to find the
ratio ($D$) between \gl~and \spc~which removed
the residual discontinuities while simultaneously
zeroing sDCA in a sample of events.
And the E-by-E correction was
modified to track the two distortions together:
\[
( \spcom + \glom ) =
( \spcrm + \glrm ) \cdot
( \dcaom / \dcarm )  \quad , \quad 
\glm \equiv D \cdot \spcm
\]

\section{Summary}
\label{sec:summary}

\begin{figure}
\begin{center}
\includegraphics*[width=67mm]{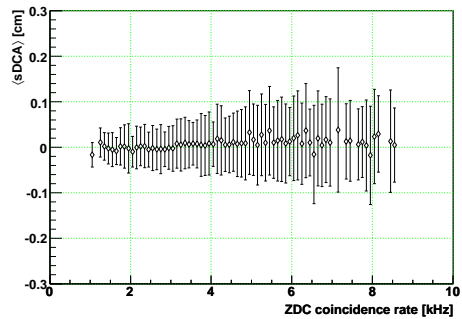}
\end{center}
\vskip -0.3cm
\caption{ \small
Performance of the ionization distortion corrections as
measured by the distributions of \dcae~(error bars are the
spread (RMS), diamonds the mean) versus luminosity (represented
by the rate of zero degree calorimeter (ZDC) coincidences~\cite{trig})
for \sne{200} GeV AuAu collisions.
} \label{fig:lum}
\end{figure}

We have identified and corrected for distortions due to ion
charge buildup
in the STAR TPC. With the onset of significant short time scale
fluctuations in
the sources of the ions which were not monitored with fine
time granularity during data acquisition, we have developed
a technique to determine and adjust for the fluctuations during
reconstruction on an event-by-event basis. Performance of the corrections
can be assessed by examining the distribution
of \dcae~as a
function of luminosity, shown in Fig.~\ref{fig:lum}. Here
we see that the spread in \dcae~is contained to within
approximately 1mm at all luminosities, and the mean is
kept to within a few hundred microns of zero.
In 2005, online monitoring with one second granularity
was implemented and will provide further assessment
of the technique's success.

\begin{thebibliography}{00}





\bibitem{TPC} 
M. Anderson \etal, Nucl. Instr. and Meth. \textbf{A499} (2003) 659.

\bibitem{HIJET}
A. Shor and R. Longacre, Phys. Lett {\bf B218} (1989) 100.

\bibitem{trig}
F.S. Bieser \etal, Nucl. Instr. and Meth. \textbf{A499} (2003) 766

\end{thebibliography}
\end{document}